# Fast processing explains the effect of sound reflection on binaural unmasking

**Norbert Kolotzek\*, Pierre G. Aublin & Bernhard U. Seeber**

Audio Information Processing, Department of Electrical and Computer Engineering, Technical University of Munich, Germany

\* Corresponding author: Norbert Kolotzek (norbert.kolotzek@tum.de)

## Abstract

Sound reflections and late reverberation alter energetic and binaural cues of a target source, thereby affecting it's detection in noise. Two experiments investigated detection of harmonic complex tones, centered around 500 Hz, in noise in a virtual room with different modifications of simulated room impulse responses (RIR). Stimuli were auralized using the SOFE's loudspeakers in anechoic space. The target was presented from the front or at 60° azimuth, while an anechoic noise masker was simultaneously presented at 0°. In the first experiment, early reflections were progressively added to the RIR and detection thresholds of the reverberant target were measured. For a frontal sound source, detection thresholds decreased while adding the first 45 ms of early reflections, whereas for a lateral sound source thresholds remained constant. In the second experiment, early reflections were cut out while late reflections were kept along with the direct sound. Results for a target at 0° show that even reflections as late as 150 ms reduce detection thresholds compared to only the direct sound. A binaural model with a sluggishness component following the computation of binaural unmasking in short windows predicts measured and literature results better than when large windows are used.

*Keywords:* binaural unmasking, binaural hearing, dynamic scenes, reverberation

**PACS:**

43.55.Hy      Subjective effects in room acoustics, speech in rooms

43.66.Dc      Masking

43.66.Nm      Phase effects

43.66.Pn      Binaural hearing

46.66.Rq      Dichotic listening

# 1 Introduction

In most real-life listening situations we are not only receiving the direct sound, but also reflections of the sound sources as multiple delayed and modified versions – in rooms and also on the street, where sound is reflected off buildings, automobiles and trees (Kuttruff, 2017). These reflections alter the interaural phase (IPD) and level differences (ILD) of the direct sound as a function of time. Such changes in the interaural cues can be helpful for detecting a target, which is based mainly two components: the better-ear monaural signal-to-noise ratio and the Binaural Masking Level Difference (BMLD) (Zurek *et al.*, 2004). Correlation changes have long been known to improve detection of a target sound in noise (Jeffress *et al.*, 1953; van de Par and Kohlrausch, 1997; 1999). Usually, the BMLD is calculated for a situation with a diotic noise masker and a dichotically out-of-phase target relative to a reference situation, where noise and target are presented diotically, as first described by Hirsh (1948). On the other hand, also a better monaural signal-to-noise ratio (SNR) at one of the ears caused by the directional dependence of the ear signals can improve detection thresholds in noise (Zurek *et al.*, 2004; Edmonds and Culling, 2006; Biberger and Ewert, 2019). Both mechanisms are frequency dependent. BMLDs, like the sensitivity to interaural phase changes, are more effective at frequencies below 1.5 kHz, whereas better-ear SNR benefits are more pronounced at higher frequencies.

Several studies investigated detection thresholds and BMLDs for different sound sources in the presence of noise and predicted the binaural benefit. van der Heijden and Trahiotis (1998) as well as Bernstein and Trahiotis (2014) measured binaural detection of a 500 Hz sine tone with opposite interaural phase ($S_\pi$) as a function of the interaural correlation of the broadband noise masker. Both studies showed that with increasing interaural correlation of the noise, binaural unmasking increases. Robinson and Jeffress (1963) measured BMLDs as a function of interaural correlation of the masker for a 500 Hz tone which was either binaurally in phase ($S_0$) or antiphasic ($S_\pi$). With increasing interaural correlation of the noise masker, BMLDs increased for a phase shifted signal and decreased for an in-phase signal. These former studies, though, did not change their interaural parameters over time.

Bernstein and Trahiotis (2017) proposed a cross-correlation model following Colburn (1977) using the mean and variance of the interaural correlation to predict the measured detection thresholds. Another model approach to predict the BMLD contribution is the equalization and cancellation (EC) theory (Durlach, 1963). Both ear signals are temporally aligned and scaled so that the interferer can be optimally cancelled. By subtracting both ear signals, the remaining energy describes the binaural benefit of the listener.

A changing interaural correlation over time, e.g. by incoming reflections, also affects the detection of a target signal in noise. Previous studies showed that for time varying interaural cues, the binaural benefit is reduced in the presence of noise, suggesting a sluggish integration process (Grantham and Wightman, 1979; Kollmeier and Gilkey, 1990; Holube *et al.*, 1998). Grantham and Wightman (1979) showed that for a sine tone in the presence of a broadband noise masker with modulated IPD, the BMLD decreases with increasing modulation frequency and becomes absent for modulation frequencies above 2 Hz. The noise masker was modulated between binaurally in-phase to binaural antiphasic. A significant reduction in unmasking was already observed for a modulation frequency of 0.5 Hz. Breebaart *et al.* (1999) also investigated the contribution of time varying interaural cues on binaural detection and proposed a model similar to the EC theory to predict measured thresholds. Their model uses the difference intensity of the left and right ear signals after peripheral processing as a detection variable. It predicts most of their data better than a cross-correlation model.

All previously mentioned models are based on the whole signal and do not evaluate changes over time in a dynamic manner. Only a few models have been proposed for signal detection in temporally changing binaural and reverberant conditions. Breebaart *et al.* (2001) proposed a binaural processing model using the same temporal resolution to extract interaural intensity and time differences to predict detection thresholds for time varying interaural conditions. Their model does not explicitly account for binaural sluggishness which is expected to influence detection thresholds of temporally changing stimuli. Braasch (2001) proposed a binaural model for detection in reverberation. It uses a 50 ms Hanning window to extract monaural and binaural contributions before both are added together, but there is no additional component taking sluggishness into account.

Binaural unmasking is seen as one contributor to binaural speech intelligibility in noise and reverberation. Beutelmann *et al.* (2010) used in their speech model the EC block proposed by Durlach (1963) along with the monaural SNR to derive the maximally possible SNR for the given interaural difference. Lavandier and Culling (2010) decomposed the binaural advantage into two separate blocks, the better ear SNR and a BMLD estimation adopted from Zurek *et al.* (2004). These models, however, estimate the binaural benefit by averaging across the whole signal or using the full room impulse response (RIR) (e.g. Rennies *et al.* (2014)) and do not specifically take into account temporal information from the incoming reflections. To consider such temporal changes, Vicente and Lavandier (2020) recently proposed a speech intelligibility model which estimates the monaural SNR benefit on short time blocks of 24 ms whereas BMLDs are derived from much longer 300 ms time window to explicitly consider a sluggish behavior of the binaural auditory system.

A coarse temporal consideration of reflections is often done for speech intelligibility, where early and late reflections are considered separately. Early reflections are commonly described to be useful whereas late reflections affect intelligibility detrimentally (Lochner and Burger, 1964; Litovsky *et al.*, 1999). Bradley (1986) found 80 ms to be the time when reflections turned from useful into detrimental for predicting the loss of speech intelligibility due to reverberation. Warzybok *et al.* (2013) measured speech reception thresholds in the presence of a single reflection for varying time delays of the reflection. They did not observe a significant difference in speech intelligibility between only the direct sound and with a single frontal reflection up to a delay of 25 ms. For larger delays they observed a moderate decrease in speech intelligibility, suggesting a partial integration of the reflection with the direct sound. Only for a delay of 200 ms, the detriment in speech intelligibility compared to only direct sound exceeded 3 dB, indicating a deteriorating effect of late frontal reflections on speech intelligibility. Nevertheless, increasing reverberation (Rennies *et al.*, 2011) or modulated noise maskers (Rennies *et al.*, 2014) remain problematic for speech intelligibility models. Although the useful-to-detrimental approach is established in speech intelligibility models, a fixed temporal boundary generic to different room acoustic conditions has been hard to find. Interestingly, little research has addressed the underlying question, what makes reflections useful or detrimental in a reverberant listening situation for binaural detection, a prerequisite for understanding speech in such situations.

In order to bridge the gap between concepts of established detection models and known speech intelligibility models, the current study deliberately goes one step back to investigate more in detail the effects of early and late reflections as well as the sluggishness integration of time varying cues in a pure detection experiment of a reverberant target signal in noise. In contrast to speech intelligibility in complex listening situations, there is no across-frequency integration in a tone-in-noise detection experiment and cognitive effects are minimized. Self-masking of speech due to the temporal smearing of phoneme information by reverberation is not relevant. With this approach, the current study focusses on the fundamental binaural concepts to better understand the perception of sound sources in reverberant situations.

To investigate the contribution of early and late reflections in a classical detection paradigm, two experiments are conducted with a harmonic complex tone with 500 Hz center frequency accompanied by simulated reflections of a room as a target signal, and an anechoic noise masker played from a single loudspeaker in the front. Experiments are conducted in an anechoic chamber and stimuli were spatially auralized via the 36 horizontal loudspeakers of the Simulated Open Field Environment (SOFE, v4) (Seeber *et al.*, 2010). To solely focus on the effect of reflections on target detection, the noise masker was anechoic. The first experiment investigates the contribution of early reflections. Detection thresholds were measured by successively adding more reflections to the direct sound. The second experiment addresses the contribution of late reflections in the same listening environment. Early reflections are successively removed from the room impulse response of the target sound. A modeling approach is investigated which evaluates the BMLD in a short time window with sluggishness taken into account later, conceptually when objects are formed. This is conceptually different to speech intelligibility models, notably Vicente and Lavandier (2020), which explicitly consider sluggishness through a slow evaluation of IPDs by using a large time frame for BMLD estimation. The proposed alternative approach with a sluggish integration only after fast extraction of the BMLDs is able to better predict the measured detection thresholds of the reverberant harmonic complex tone in noise.

## 2    First experiment: Contribution of early reflections to binaural unmasking

### 2.1    *Experimental setup*

Experiments were conducted in the Simulated Open Field Environment (SOFE, v4) (Seeber *et al.*, 2010) in the anechoic chamber at the Technical University of Munich. The stimuli were presented via the SOFE's 36 horizontally arranged loudspeakers (Dynaudio BM6A mkII, Dynaudio, Skanderborg, Denmark) placed in 10°-spacing. The loudspeakers are mounted on a custom 4.8 m x 4.8 m squared holding frame in a height of 1.4 m. The loudspeaker at 0°, in front of the listener, has a distance of 2.4 m to the listener's position. Loudspeaker-individual finite-impulse response equalization filters of length 512 taps (at $f_s$=44100, time-shifted in a 1024 taps filter) were used during playback to compensate for the frequency and phase response and the time-of-arrival difference.

## 2.2 Simulated room configuration

A non-rectangular virtual room was simulated with two different absorption coefficients $\alpha_1 = 0.1$ and $\alpha_2 = 0.5$. Figure 1 illustrates the virtual room including the simulated listener position and the two simulated source positions at 0° and 60° at a distance of 5 m from the virtual listener position. Direct-to-reverberant ratios were derived for the 0° and 60° source position to -11.8 dB and -12.3 dB, respectively, for $\alpha_1 = 0.1$, and to -4.2 dB and -4.9 dB for $\alpha_2 = 0.5$. The reverberation time $RT_{60}$, was 736 ms and 302 ms for $\alpha_1$ and $\alpha_2$, respectively. In the room simulation, only specular reflections were simulated. To avoid standing waves and strictly repetitive reflection times, the room corners were shifted by up to 50 cm from a rectangular configuration, which results in a somewhat natural temporal jittering of the room reflections. The exact corner coordinates are listed in Table A2 of the appendix.

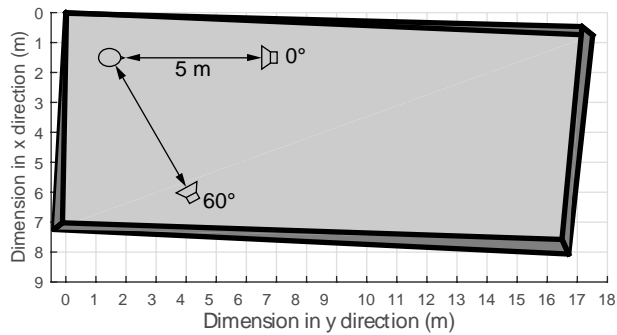

Figure 1: Sketch of the simulated room in topview with source and receiver positions. The room is 3 meters high. All room corners were shifted by up to 50 cm to prevent strictly parallel walls in the room. This avoids standing waves and strictly repetitive reflection times, and thus introduces a more natural temporal jittering of the room reflections. The receiver was placed in the top left corner with a distance of 1.5 meters from the walls and at a height of 1.4 meters, corresponding to approximately the seating height of a person. Both simulated sound sources had a fixed distance of 5 meters to the listener's position. The condition with the frontal target will be denoted as $S_0$, the one at 60° as $S_{60}$.

All surfaces of the room were covered with the same theoretical material, having either an absorption coefficient of 0.1 or 0.5 for each octave frequency band from 125 Hz to 4 kHz. Room impulse responses (RIRs) were generated using the SOFE (Seeber *et al.*, 2010), which is based on the image source method. Specular reflections were simulated up to 100$^{th}$ order while all image sources with more than 7 invisible parents in a row or a level 80 dB below the direct sound were ignored. For the first experiment, RIRs for two absorption coefficients ($\alpha_1 = 0.1$ & $\alpha_2 = 0.5$), two source positions (0° & 60°) were generated. To test the effect of early reflections, the RIRs were truncated after 15 ms (only direct sound), 20 ms, 45 ms, 75 ms, 150 ms, 250 ms, and 500 ms.

To reproduce the simulated room over the SOFE's 36 horizontally arranged loudspeakers, 17$^{th}$-order Ambisonics was used according to the encoding in Zotter and Frank (2019, p. 61, Eq.4.19) with *max r$_E$* decoding (Daniel, 2001) to maximize the energy vector $\vec{r_e}$ of the sound field. This results in 36 room impulse responses for each loudspeaker per tested condition.

## 2.3 Stimuli

Since BMLDs are known to be more salient at frequencies below 1.5 kHz, a harmonic complex tone (HCT) consisting of the 7$^{th}$ to 13$^{th}$ harmonic to 50 Hz fundamental frequency (350 Hz to 650 Hz) was used to generate the target stimulus centered around 500 Hz. Since for truly resolved harmonics reflections will only affect each harmonic's energy and phase, this HCT with unresolved harmonics was chosen to provide envelope fluctuations. The level of each harmonic was set such that each auditory filter, with a width defined on the Bark scale, received identical energy. The target stimulus was convolved with the truncated room impulse responses for each of the 36 loudspeakers, resulting in 36 loudspeaker signals. The level at the listener's position (sum across all loudspeaker channels) of the reverberant signals was then normalized across different truncation conditions. The reverberant HCT had an effective duration of 500 ms, defined as the envelope exceeding 90% of its maximum (Kolotzek and Seeber, submitted), with 10 ms Gaussian rise and fall times.

Uniform exciting noise was used as masker (Fastl and Zwicker, 2007). The noise was band-limited from 250 Hz to 750 Hz, to ensure masking all components of the HCT target stimulus without becoming too loud. It had an overall duration of 900 ms with 30 ms Gaussian rise and fall times. The noise source had a sound pressure level of 60 dB at the listener's position. The noise was chosen to be anechoic and not filtered with the room impulse response to avoid interaction by reflections of the noise masker. It was played from a single loudspeaker at 0°, in front of the listener, leading to binaurally highly correlated noise with an interaural correlation coefficient of 0.99. The correlation coefficient was determined from binaural recordings at the listener's position with the HMS II.3 artificial head with an anatomically formed pinna (Type 3.3) according to ITU-T P.57 (HMS II, Head acoustics GmbH, Herzogenrath, Germany).

## 2.4 Participants

Eight participants (3 female) volunteered for the experiment. Participant's age ranged from 21 to 29 years (mean: 25 yr.; sd: 2.3). All participants had normal hearing thresholds with a hearing loss less than 15 dB up to 8 kHz as assessed with a clinical audiometer (Madsen Astera2, GN Otometrics A/S, Taastrup, Denmark). All participants gave written consent and were not payed for participating in the experiment. The study was approved by the ethics committee of the TUM, 65/18S.

## 2.5 Procedure

The participants sat in the completely darkened anechoic chamber in the center of the loudspeaker array. The detection threshold of the HCT in noise was determined with a three-interval three-alternative-forced-choice method (3I-3AFC) using a two-down/one-up adaptive staircase procedure (Levitt, 1971) tracking the 71% point of the psychometric function, similar to Kolotzek and Seeber (submitted). Participants listened to three intervals of the anechoic uniform exciting bandpass noise, separated by an inter-stimulus-interval of 500 ms. To one of these intervals the reverberant target HCT was added. After the stimulus presentation (3.7 s duration), the listeners' task was to indicate which interval differs from the others by pressing the corresponding number on a keyboard. Depending on their response, the overall level of the HCT was adjusted. The initial level was set to 65 dB SPL at the listeners' position with an initial step size of 5 dB. After the first reversal, the step size was decreased to 2 dB. From the fourth reversal onwards it was further decreased to the final step size of 1 dB. Twelve reversals were measured at the final step size and the mean of the last ten reversals was used to calculate the detection threshold of the HCT in noise.

The experiment was blocked by the absorption coefficient $\alpha$. The order of the blocks was randomized between subjects. The combination of used RIR truncation time and target location was randomized within each block. Before a new random test condition started, the previous one had to be finished (blocked by track), i.e. tracks were not interleaved to avoid potential issues with spatial attention due to the changing target location. Each subject finished one track for each condition, resulting in 28 tracks for each subject. Subjects finished the experiment on average in 2 hours.

## 2.6 Results

Medians with quartiles of the measured thresholds for both absorption coefficients ($\alpha_1$ = 0.1 & $\alpha_2$ = 0.5) and both source positions (0° & 60°) are shown in Figure 2. For a sound source positioned at 0° in front of the listener, thresholds decrease with an increasing amount of reflections, which suggests that adding early reflections helps to detect the HCT from the front in noise. A similar behavior can be seen for both absorption coefficients. Even when adding only a few early reflections (e.g. truncation after 20 ms), thresholds decrease by more than 5 dB compared to only the direct sound (truncation after 15 ms). Interestingly, such an improvement with increasing number of reflections cannot be observed for a target sound source at 60°. Here, thresholds are 15 dB lower for only the direct sound compared to a target positioned at 0°, because of spatial masking release. When adding early reflections, there is no additional benefit. A slight negative effect can be observed when adding reflections later than 150 ms. Here, thresholds for both absorption conditions increase by 1 to 2 dB and a similar behavior for both absorption coefficients can be observed.

Repeated measures analysis of variance (rmANOVA) with target position, absorption coefficient and truncation as within-subjects variables was performed on the measured data. The main effects of target position [$F(1,7) = 1027$, $p<0.001$] and truncation [$F(6,42) = 152$, $p<0.001$], and the two-way interactions of position and truncation [$F(6,42) = 113$, $p<0.001$] and of position and absorption [$F(1,7) = 13$, $p<0.01$] and the three-way interaction [$F(6,42) = 2.7$, $p<0.05$] are significant. Since there is no main effect of the absorption coefficient and only the two-way interaction of position and absorption is significant, but not the interaction of absorption and truncation, this indicates that the different absorption coefficients do not affect the binaural benefit as seen in similar results for both absorption conditions. The significant interaction of absorption and position can be explained by the difference in thresholds for short truncations between the two different target positions (see Fig. 2 solid versus dashed lines for truncation 15 ms to 45 ms). To further analyse the interactions, a two-tailed t-test post-hoc analysis with Tuckey-Kramer correction was performed. For a sound source at 60°, no pairwise comparison of the different truncation times reaches significance for both absorption coefficients, which suggests that there is no further unmasking benefit from the reflections for a lateral target position.

For the target position at 0°, Tuckey-Kramer corrected two-tailed t-test pairwise comparisons show a significant difference between 15 ms and all other truncation times (p<0.001), between 20 ms and all other truncation times (p<0.05) and between 45 ms and 150 ms (p<0.05). No other combination reaches significance. This indicates that the binaural benefit from adding early reflection for a target position at 0° increases up to a truncation time of about 45 ms. Adding later reflections after 150 ms will not further improve the detection of the target HCT in noise from the front.

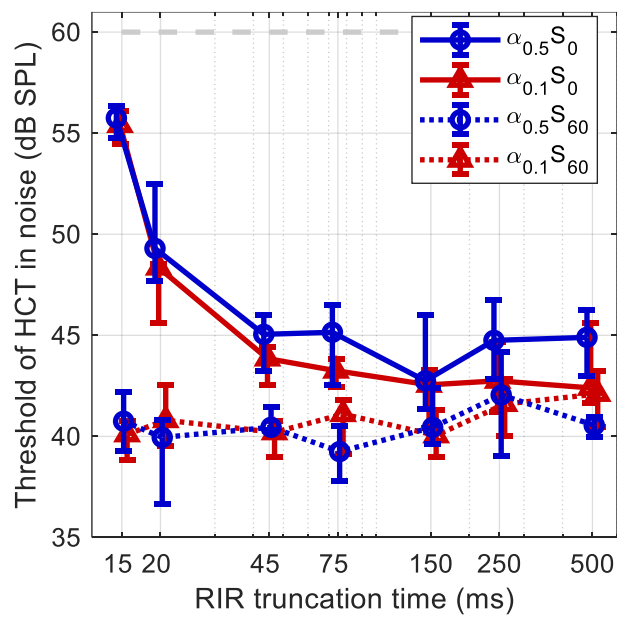

Figure 2: Measured binaural thresholds of a reverberant harmonic complex tone (HCT) for different truncations of the RIR in the presence of an anechoic bandpass noise (BPN) with 60 dB SPL from the front are shown. With increasing RIR truncation time, the amount of late reflections increases, while a 15 ms window corresponds to only the direct sound without any reflections. Solid lines indicate thresholds for a source at 0°, dashed lines for a source at 60°. Blue circles show the median thresholds of the tested participants for an absorption coefficient of 0.5 and red triangles for an absorption coefficient of 0.1. Errors are given as upper and lower quartiles.

## 3   Second experiment: Unmasking in the absence of early reflections

The aim of the second experiment is to focus on the effect of only late reflections on binaural unmasking of a target sound source in noise. It was shown for speech intelligibility that late reflections can harm the intelligibility (Lochner and Burger, 1964; Lavandier and Culling, 2010; Warzybok *et al.*, 2013). These studies found that reflections arriving within the first 80 to 100 ms after the direct sound can be integrated with the direct sound, whereas later reflections will not contribute to intelligibility and can be interpreted as being energetically added to the masking background noise. The main question in this experiment is whether late reflections will also hinder the simple detection of a reverberant target tone in the presence of noise and if also here late reflections will add additional energy to the masking signal. The experiment was similar to the first one, but early reflections were increasingly removed from the RIR and late reflections were kept along with the direct sound.

### *3.1 Stimuli*

The second experiment used the same room and absorption coefficients, but only the target source position at 0° in front of the listener since there was no change in threshold for a source positioned at 60°. In contrast to the first experiment, early reflections were removed from the RIR so that, besides the direct sound, only reflections after a certain time were kept. These times correspond to the same truncation times as in experiment 1 (15 ms, 20 ms, 45 ms, 75 ms, 150 ms, 250 ms, and 500 ms) with all reflections between the direct sound and the truncation time being removed from the RIR. Therefore, 500 ms correspond to only the direct sound, whereas 15 ms in this case corresponds to the full impulse response. The longer the cutting time condition, the larger the gap between direct sound and incoming reflections. To illustrate the different modifications of the RIR, Figure 3 shows schematically the difference of the cut RIRs used in the first and in the second experiment.

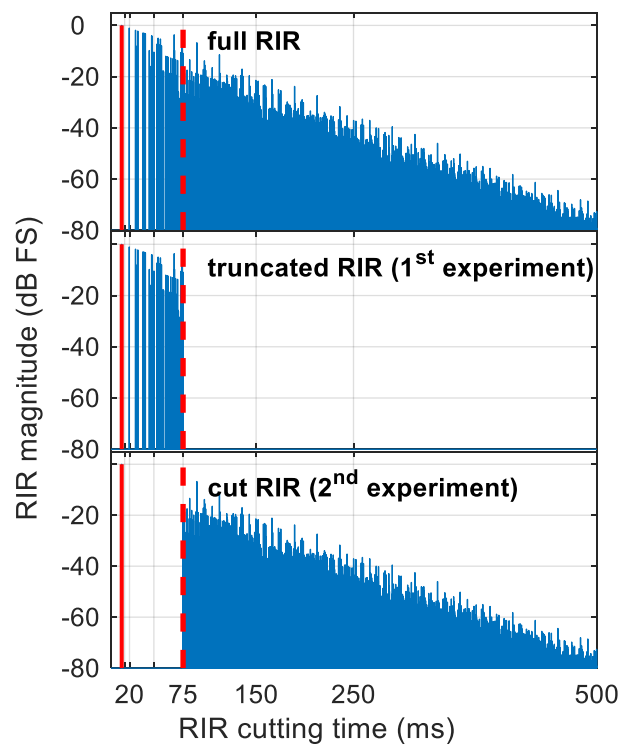

Figure 3: Schematic representation a full RIR (top row) and the modified RIRs of the first experiment (2nd row) with truncated RIR after 75 ms and the cut RIR of the second experiment (bottom row) where reflections where zero'ed after the direct sound to the time condition given. In all experimental conditions the direct sound is always preserved in the RIR.

The same HCT target stimulus as in experiment 1 was convolved with the cut impulse responses and the level was normalized across different time conditions. The noise masker had the same frequency range and duration as in experiment 1.

### 3.2 Procedure

The same eight volunteers also finished the second experiment in about 1 hour. The experimental procedure followed that of experiment 1. Trials were blocked by the absorption coefficient and randomized between subjects. Within each block, RIR truncations were randomized, but tracks were not interleaved. Each subject finished one track for each condition, resulting in 14 tracks for each subject.

### 3.3 Results

Thresholds obtained from the second experiment are summarized in Figure 4. Removing the very first early reflections does not seem to have an impact on the thresholds, as they remain fairly constant between 15 ms and 20 ms truncation time for both $\alpha$. However, as more and more early reflections are removed, thresholds start to increase from 45 ms to 150 ms for $\alpha = 0.5$, and stay constant thereafter on the same level reached by only the direct sound (500 ms). For $\alpha = 0.1$, a slightly different behavior can be observed. Thresholds for 45 ms truncation time decrease first and start to increase for truncation times larger than 150 ms. Different to the first experiment, absorption influences measured thresholds as it determines the truncation time from which thresholds start to increase.

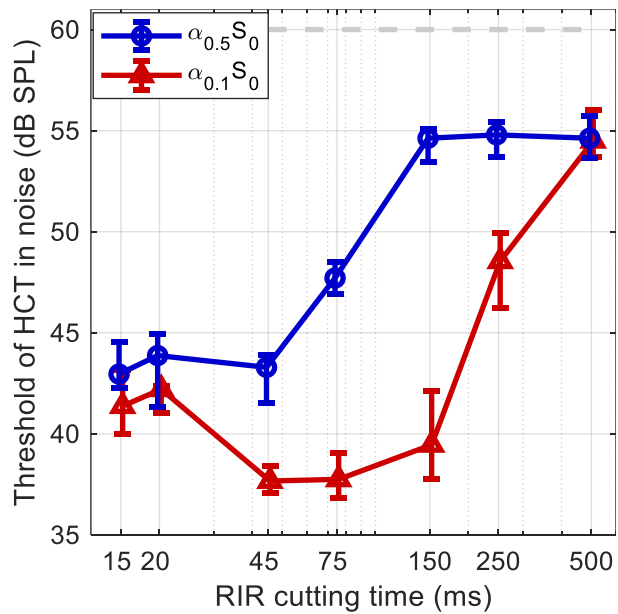

Figure 4: Measured binaural thresholds of a reverberant harmonic complex tone (HCT) for different time conditions of cut early reflections from the RIR in the presence of an anechoic noise with 60 dB SPL. Both sound sources were colocated at 0°. The blue circles show median thresholds of the tested participants for an absorption coefficient of 0.5, the red triangles for an absorption coefficient of 0.1. Errors are given as upper and lower quartiles.

An rmANOVA with absorption coefficient and truncation time as within-subject variables was performed on the measured thresholds. Besides a significant main effect of truncation time [$F(6,42) = 185$, $p<0.001$] also the main effect of absorption coefficient [$F(1,7) = 334$, $p<0,001$] and the interaction between truncation time and absorption coefficient [$F(6,42) = 56$, $p<0.001$] become significant. In a post hoc analysis with Tuckey-Kramer correction, pairwise comparison between $\alpha = 0.1$ and $\alpha = 0.5$ shows no significant difference for only direct sound (500 ms) and for 20 ms. All other truncation times are significantly different between absorption coefficients (p<0.05 for 15 ms, else p<0.001) which suggests that late reflections in more reverberant situations (α = 0.1) strongly affect detection thresholds. A pairwise comparison of the measured thresholds between different truncation times shows no significant difference by removing the very first reflections for both absorption coefficients (15 ms vs. 20 ms for α = 0.1 and 15 ms vs. 20 ms & 45 ms for α = 0.5). The decrease in detection thresholds observed for α = 0.1 between 20 ms and 45 ms is significant (p<0.001). For an absorption coefficient of 0.5, very late reflections (truncation times larger than 150 ms) do not change detection thresholds compared to only the direct sound, since thresholds are not significantly different from each other.

## 4 Short vs long window binaural processing for detection of reverberant signals

### 4.1 Short window, fast binaural processing model (BMLD*fast*)

Starting point of the current approach for a binaural processing with fast BMLD formation (BMLD_fast) was a simplified version of the model proposed by Lavandier and Culling (2010), published in the Auditory Modelling Toolbox (AMT) (Søndergaard and Majdak, 2013). In the model, the overall binaural benefit is divided into two parts, better ear SNR and the BMLD. Both parts are extracted for each critical band separately. The EC formula used in the current model approach was adopted from Lavandier and Culling (2010) and is shown in Equation 1, where $f_i$ denotes the center frequency of a particular auditory filter, $\phi_T$ is the interaural phase difference of the target, $\phi_M$ is the interaural phase difference of the noise masker and $\rho_M$ denotes the interaural coherence of the noise masker.

$$BMLD(f_i) = \max\left(k - \frac{\cos(\phi_T(f_i) - \phi_M(f_i))}{k(f_i) - \rho_M(f_i)}\right) \quad (1)$$

$k(f_i)$ can be derived by $k(f_i) = (1 + \sigma_\epsilon^2) \exp\left((2\pi f_i)^2 \sigma_\delta^2\right)$ according to the formula given in Lavandier and Culling (2010), with $\sigma_\epsilon = 0.25$ and $\sigma_\delta = 0.105 \cdot 10^{-3}$ (Durlach, 1972).

The overall structure of the BMLD_fast model approach is shown in Figure 5. Both, noise and target signal, are filtered with a Gammatone filter bank, according to the Bark scale, separately for the left and the right ear. The output of the Gammatone filter bank is then split into 24 ms time frames using a Hanning window with 50% overlap of successive time frames. The effective window length of the Hanning window is therefore 12 ms measured by exceeding -6 dB of its maximum. For each frequency band and time window, the interaural phase difference of the target and the masker noise as well as the interaural coherence of the noise masker are derived using the interaural cross correlation. The extracted interaural cues are used to compute the BMLD according to Equation 1, for each auditory filter and short time window. The main difference to former models is that the BMLD contribution is derived on short time windows before taking sluggishness into account. In the BMLD_fast approach, only after formation of the BMLD contribution on short time blocks a 300 ms exponential decay filter is applied to the BMLD time series, to account for the sluggishness of the auditory system. The exponential decay was used to weight the onset of incoming cues more strongly and to introduce a biologically inspired forgetting factor. Thereafter, the BMLD contribution is transformed to decibels (Lavandier and Culling, 2010).

In addition to the BMLD, the better ear SNR was derived from the binaural ear signals. Similar to the processing of the BMLDs, the signal-to-noise ratios for both ear signals were computed separately in each short time frame and for each auditory filter. The better SNR across both ears signals is chosen. To account for temporal integration, the better ear SNR is filtered with a 200 ms exponential integration filter (Fastl and Zwicker, 2007). After that, the SNR is transformed into level given in decibels. Both BMLD and better ear SNR are summed for each time frame and for each critical band, resulting in an overall binaural benefit. To model a simple detection process, the model selects the frequency band with the highest overall binaural benefit in each time frame followed by selecting the maximum of the time series.

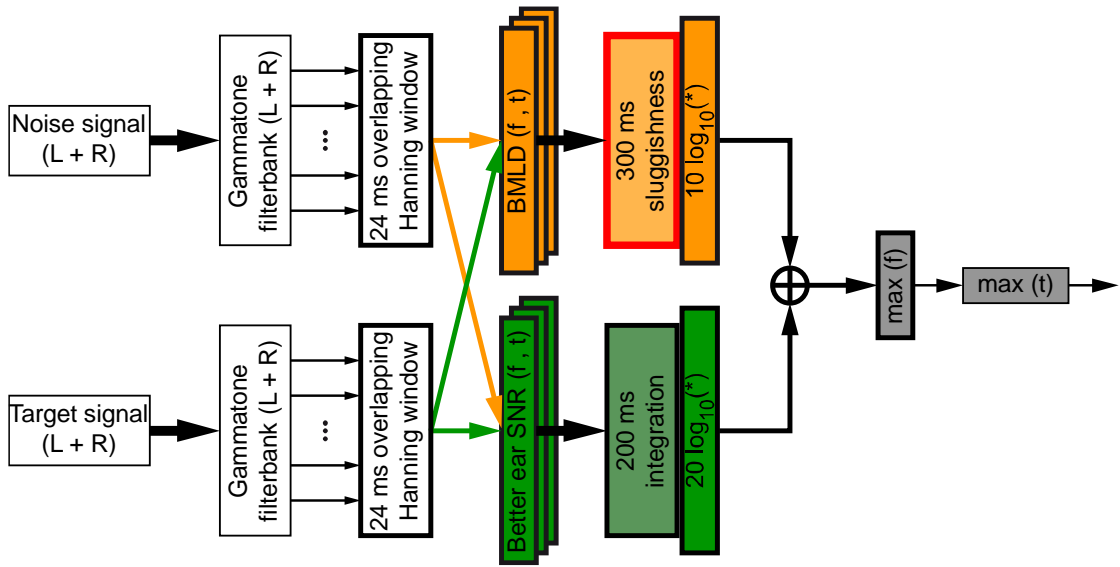

Figure 5: Scheme of the short-window, fast processing approach (BMLD$_{fast}$). The left and right ear signals are first bandpass filtered using a Gammatone filter bank parametrized along the Bark scale. The time signal of each filter is windowed with 24 ms overlapping Hanning windows resulting in an effective window length of 12 ms. The interaural cross-correlation of the interferer ($\rho_i$) as well as the interaural phase difference of target and interferer ($\phi_t$ & $\phi_i$) are extracted for each filter and each time window to calculate binaural unmasking according to formula 1. A 300 ms exponential decay filter is subsequently used to account for sluggishness of binaural processing. Binaural unmasking and the better ear SNR are added for each frequency band and time frame, followed by selecting the maximum of the binaural benefit across frequency bands per time frame. The binaural benefit for signal detection is estimated by selecting the maximum binaural benefit over time. The BMLD$_{slow}$ approach differs from BMLD$_{fast}$ only by using a 300 ms window directly after the Gammatone filterbank of derive the BMLDs without integration afterwards, else both models are identical.

### *4.2 Long-window, slow binaural processing approach (BMLD$_{slow}$)*

The BMLD$_{fast}$ approach is compared to a slow binaural processing model (BMLD$_{slow}$), which differs only in a few details. BMLD$_{slow}$ uses two different time frames, a fast 24 ms frame to extract the better ear SNR identical to the BMLD$_{fast}$ approach, and a 300 ms frame to compute the BMLD contribution directly after the Gammatone filter bank. Since the temporal integration is thus done on signal level in the longer BMLD frame, no additional sluggishness filter is applied after extracting the BMLD. The final prediction is made identical to the BMLD$_{fast}$ approach.

### *4.3 Evaluation*

#### 4.3.1 Current experimental conditions

Both model approaches were evaluated with the current psychoacoustical results. In-situ binaural recordings were used as input signals for the model. The in-situ signals of the anechoic noise masker and of the reverberant target were recorded with an artificial head at the listener's position in the SOFE (see methods). The model predictions for all tested RIR conditions and source positions are shown together with the experimental results in Figure 6.

Predictions with the BMLD$_{fast}$ approach follow the measured data well across most conditions. The data from the first experiment with collocated target and masker at 0° (panel a & b) can be predicted well with the fast BMLD extraction. The root mean square error (RMSE) of the predictions to the experimental data is 0.98 dB and 1.55 dB for an absorption coefficient of 0.1 and 0.5, respectively. The Pearson's correlation coefficient expresses a high correlation ($\rho = 0.98$; $\rho = 0.94$) for both absorption coefficients. With the BMLD$_{slow}$ approach, the overall binaural benefit is underestimated for an absorption coefficient of 0.5. This can also be seen in the high RMSE of 3.16 dB, which is twice as large as for the BMLD$_{fast}$ approach in this condition. Also for an absorption coefficient of 0.1, the RMSE is with 1.38 dB higher for the BMLD$_{slow}$ approach compared to BMLD$_{fast}$. The correlation for BMLD$_{slow}$ vs the data is nevertheless high for both absorption coefficients ($\rho = 0.95$ for $\alpha = 0.1$; $\rho = 0.98$ for $\alpha = 0.5$).

Predictions for the N0S0 condition also differ between fast and slow BMLD formation for the second experiment (panel e & f). When early reflections are successively cut out, the difference between a sluggish integration before or after the formation of the BMLD contribution is clearly visible for truncation times larger than 75 ms, especially for higher reverberation ($\alpha = 0.1$). Here, BMLD$_{slow}$ leads to an underestimation of the measured thresholds whereas BMLD$_{fast}$ matches the measured thresholds well. This can also be observed in the RMSE and the correlation of the predictions to the measured data. While RMSE is 0.98 dB and 2.12 dB for the BMLD$_{fast}$ approach, errors increase for the BMLD$_{slow}$ approach to 4.39 dB and 3.54 dB for $\alpha = 0.1$ and $\alpha = 0.5$, respectively. This is mainly because of the huge underestimation of unmasking for late incoming reflections in the BMLD$_{slow}$ approach. With BMLD$_{fast}$ predictions are highly correlated with the measured threshold data ($\rho = 0.99$) for both absorption conditions, whereas with the BMLD$_{slow}$ approach correlation decreases to 0.83. One reason for the better performance with BMLD$_{fast}$ is that faster interaural correlation changes, caused by late incoming reflections, are smeared over time when using a longer time window for BMLD estimation. Such fluctuations in the interaural correlation will be kept with a fast estimation of the BMLD. For a target sound source located at 60° for a frontal noise masker (panel c & d) the overall performance of both model approaches does not differ much. The RMSE is 2.62 dB and 1.93 dB for BMLD$_{fast}$ and 2.98 dB and 2.19 dB for BMLD$_{slow}$ for $\alpha = 0.1$ and $\alpha = 0.5$ respectively. Pearson's correlation coefficients are quite low for an azimuth condition of 60° and stay in the range of 0.17 to 0.2 for both model approaches. The low $\rho$ values here can be explained by considering that across truncation time there is no change that can be predicted. RMSE as well as Pearson's correlation coefficients are summarized in Table 1 for both experiments and model approaches.

Table 1: Root mean square errors (RMSE) and correlation coefficients ($\rho$) of the $BMLD_{fast}$ and $BMLD_{slow}$ predictions to the experimental data for all tested conditions.

|  | **$RMSE_{fast}$** | $\rho_{fast}$ | **$RMSE_{slow}$** | $\rho_{slow}$ |
|---|---|---|---|---|
| **Exp. 1, $S_0$ $\alpha = 0.1$ early reflections** | 0.98 dB | 0.98 | 1.38 dB | 0.95 |
| **Exp. 1, $S_0$ $\alpha = 0.5$ early reflections** | 1.55 dB | 0.94 | 3.16 dB | 0.98 |
| **Exp. 1, $S_{60}$ $\alpha = 0.1$ early reflections** | 2.62 dB | 0.17 | 2.98 dB | 0.18 |
| **Exp. 1, $S_{60}$ $\alpha = 0.5$ early reflections** | 1.93 dB | 0.17 | 2.19 dB | 0.20 |
| **Exp. 2: $S_0$ $\alpha = 0.1$ late reflections** | 0.92 dB | 0.99 | 4.39 dB | 0.83 |
| **Exp. 2: $S_0$ $\alpha = 0.5$ late reflections** | 2.12 dB | 0.99 | 3.54 dB | 0.83 |

The overall trend and most of the tested conditions can be predicted quite well. The overall average error of the model predictions to the measured data across all tested conditions is 1.8 dB for the BMLDfast approach and 3.1 dB for the BMLDslow approach. Some conditions, though, cause difficulties for both approaches: in panel c & d at 15 and 20 ms and in panel f at 20 and 45ms cutting time. Adding only very early reflections to a lateral sound source or cutting out early reflections from a frontal sound source results in an overestimation of the overall binaural benefit in both model approaches. This is likely caused by the better-ear SNR contribution since it is much higher than for all other tested truncation times, whereas the BMLD contribution is constant across tested truncation times. The difference to the measured data reaches here a maximum of 4.6 dB (panel c at 20 ms truncation time).

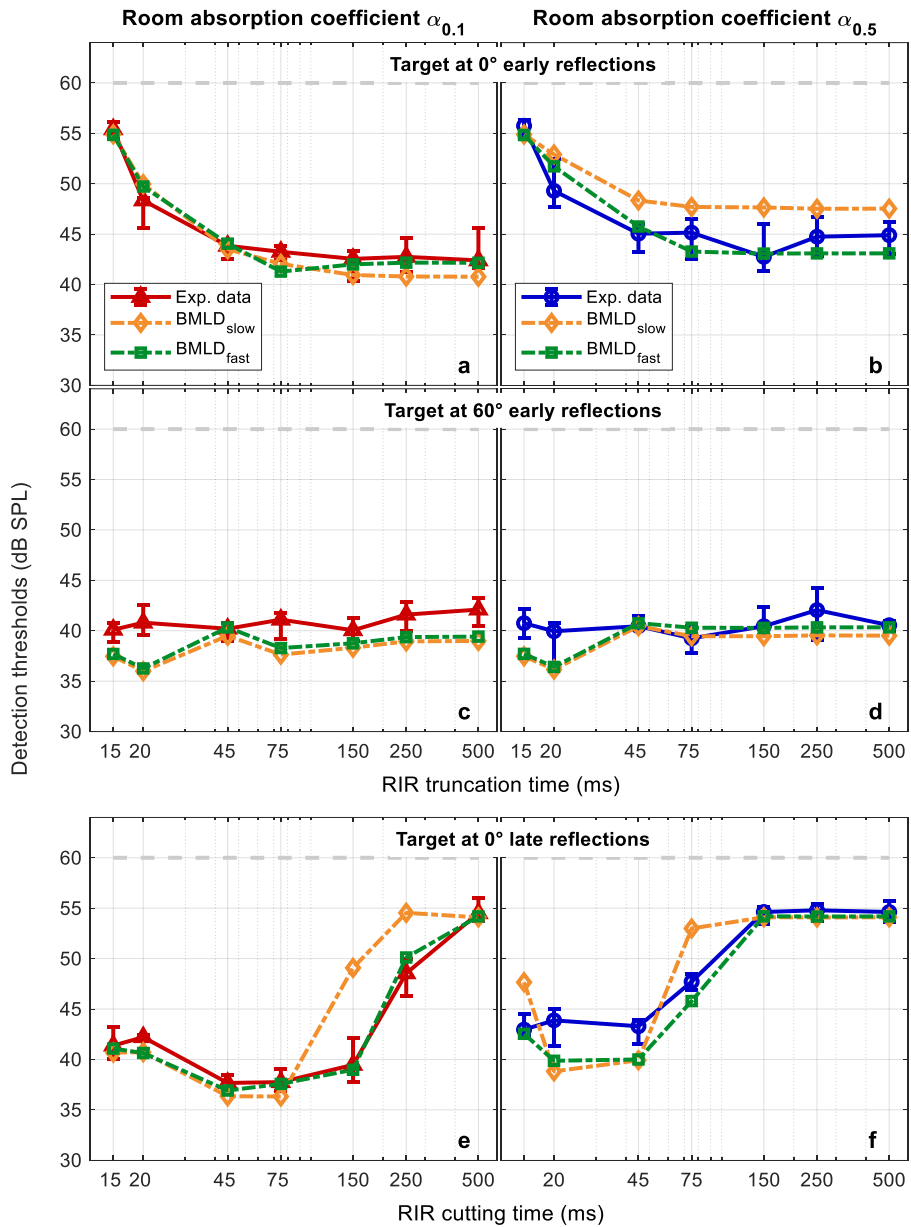

Figure 6: Predictions of the short-window model approach (BMLD$_{fast}$) (green dashed lines with squares) with a sluggishness integration of short-time BMLDs are shown along with predictions using a 300 ms window for BMLD extraction (BMLD$_{slow}$) (orange dashed lines with diamonds) against measured results. Data are given as a function of cutting time of the room impulse response. The left column (a, c & e) shows predictions for an absorption coefficient of $\alpha = 0.1$ and the right column (b, d & f) for $\alpha = 0.5$. The first row (a & b) shows the prediction for sound source and noise being co-located in the front of the listener (N$_0$S$_0$), the second row (pannel c & d) for a sound source at 60° (N$_0$S$_{60}$). The third row (e & f) shows predictions for the second experiment (N$_0$S$_0$ with only late reflections). The experimentally measured binaural unmasking (solid lines) are replotted from Figure 2: (pannel a – d) and Figure 4: (pannel e & f) for comparison.

### 4.3.2 Evaluation on binaural detection experiments in the literature

To further evaluate the differences between both, the $BMLD_{fast}$ and $BMLD_{slow}$ approaches, two additional data sets from the literature were used. Braasch (2001) measured detection thresholds of a reverberant broadband noise signal at different azimuth angles, 0°, 2° and 20°. Another broadband noise was used as masker located at 0°. Both noises, target and masker, had a frequency range of 200 Hz to 14 kHz and were presented from a distance of 2 m to the virtual listener position. A rectangular room (5m x 6m x 3m) was simulated using the mirror image technique (Allen and Berkley, 1979), but reflections formed temporally repetitive patterns. Measured thresholds of stimuli with all binaural cues available are replotted from Fig 5.9 in Braasch (2001) and are shown with the predictions in Figure 7 in the left graph. The thresholds predicted with the $BMLD_{fast}$ approach match the measured data of Braasch (2001) for almost all conditions. Only for a target source located at 0°, thresholds are underestimated by approximately 2.5 dB, just outside the across-subject variance. The RMSE of the predicted benefit against the provided measured data is 1.75 dB. The figure also shows predictions of the $BMLD_{slow}$ approach. The overall decrease of the binaural benefit with increasing azimuth angle can also be predicted, but BMLDs are overall underestimated, resulting in an RMSE of 4.48 dB.

The second data set for comparing both model approaches is taken from a study by Zurek *et al.* (2004). The room simulated in this study was also rectangular (4.8 m x 6.6 m x 2.6 m), with the virtual listener placed near its middle, 2.8 m from the right wall and 2.5 m from the rear wall. The listener was turned by 20° to the left. They used a $3^{rd}$-octave bandpass noise with a center frequency at 500 Hz as target stimulus and a continuous broadband noise as masker. Detection thresholds of the reverberant target at 0° in 1 m distance to the listener were measured in an anechoic noise masker at 60° azimuth and 1 m distance for different absorption coefficients. Binaural room impulse responses were derived with a spherical head model with 8.75 cm head radius. Their threshold data, relative to averaged thresholds measured only presenting to the left or right ear, are replotted from Fig. 7e in Zurek *et al.* (2004) and are shown with the model predictions in the right panel of Figure 7. The $BMLD_{fast}$ approach predicts their results across all tested absorption coefficients well with a slight underestimation of the binaural benefit resulting in an overall RMSE of 1.45 dB. The $BMLD_{slow}$ approach captures the trend of a decreasing binaural benefit with increasing absorption coefficient, but errors increases with more reverberation (RMSE = 3.12 dB). Results indicate that using fast BMLD extraction followed by sluggish integration is beneficial for prediction of highly reverberant conditions. This is also in line with results from the current study, showing that the $BMLD_{fast}$ approach predicts the benefit caused by late reflection in highly reverberant situations better.

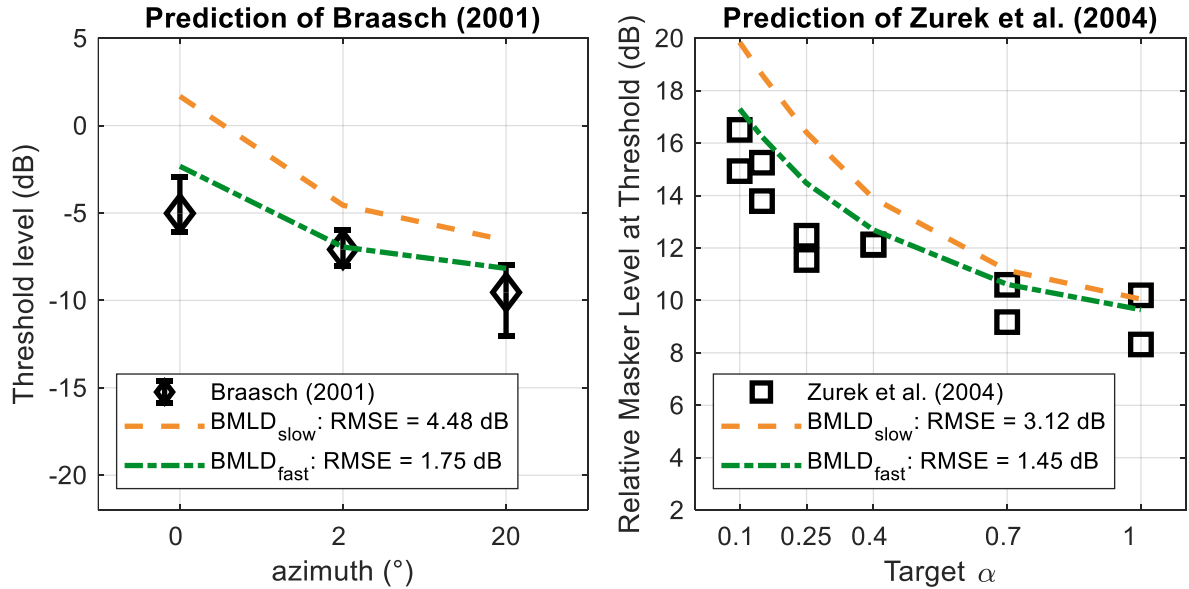

Figure 7: Predictions of the BMLD$_{fast}$ (green dashed-dotted line) and BMLD$_{slow}$ (orange dashed line) model approaches for two former detection experiments in reverberent environments are shown. The left panel shows predictions for detection data by Braasch (2001) for a reverberant broadband noise target at different azimuth positions in the presence of another broadband noise masker at 0°. The right panel presents data of two subjects collected by Zurek *et al.* (2004) for a 3$^{rd}$-octave bandpass noise target at 0° in reverberant space with different absorption coefficients when an anechoic broadband noise masker was presented from +60°.

## 5 General Discussion

This study investigates how early and late room reflections affect the detection of a harmonic complex tone in the presence of a noise masker in realistic free-field listening conditions. Two experiments were conducted in a room simulated with two different absorption coefficients. Listeners detected a reverberant harmonic complex tone, centered around 500 Hz and located at 0° or 60°, in an anechoic uniform-exiting noise masker presented from the front. The first experiment focused on the effect of early reflections on detection by subsequently adding reflections to the direct sound of the target, whereas the second experiment investigated the influence of late reflections by subsequently cutting out early reflections from the full room impulse response. Two modelling approaches are compared, one approach where interaural cues for BMLD computation are extracted on a larger time frame (300 ms; BMLD$_{slow}$), and a suggestion for a dynamic approach operating on short time frames for BMLD computation with binaural sluggishness taken into account only afterwards (BMLD$_{fast}$). The BMLD$_{fast}$ approach excels when predicting thresholds of a reverberant harmonic complex tone in noise presented from the front for various literature data over the BMLD$_{slow}$ approach. The results suggest that a fast extraction of the binaural benefit with sluggishness applied only afterwards matches detection thresholds more precisely than a slow extraction of BMLDs, especially in higher reverberation and non-standard situations with only late reflections.

### 5.1 Effects of early reflections on signal detection in noise

Results of the first experiment of this study show that early reflections improve detection thresholds of a low frequency harmonic complex tone in static noise if the target sound source is collocated with the masker at 0° in front of the listener. In this condition, the direct sound does not provide advantageous binaural information to unmask the target signal (comparable with an $N_0S_0$ condition in a classical BMLD experiment). Adding early reflections up to 75 ms decreases the interaural correlation of the target which results in an increased binaural benefit. Adding later reflections does not further decrease the interaural correlation, which might explain the constant thresholds obtained when adding additional reflections after 75 ms. To illustrate these observations, Figure 8 shows the time course of the interaural correlation (IC) of the reverberant target signal located at 0° for an absorption coefficient of 0.1 for different RIR truncations.

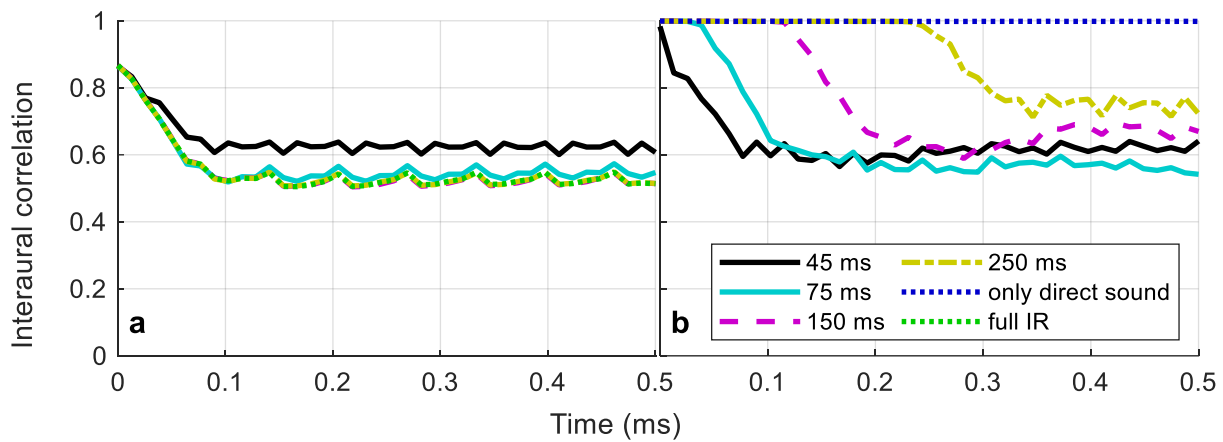

Figure 8: Short-term interaural correlation (IC) of the reverberant target signal located at 0° depending on the time point of the signal. The parameter varied between curves is the cutting time condition of the RIR, in panel a) given as truncation time (experiment 1), in panel b) the RIR was zero'ed after the direct sound to the time condition given (experiment 2). When adding early refections up to a truncation time of 75 ms (panel a), the IC decreases, and it remains at a constant, low value when later reflections are added. Late reflections decorrelate the fronal target signal, but the maximum decorrelation requires reflections to arrive within 150 ms (panel b).

Zurek *et al.* (2004) measured detection thresholds of a 1/3 octave narrowband noise with a broadband noise masker in simulated reverberation. Monaural thresholds in the anechoic condition were compared to binaural thresholds in reverberation. Their results for collocated target and masker at 0° suggest that reverberation does not have a significant impact on detection thresholds. This is in contrast to the results of the current study, which clearly show that adding early reflections to a frontal target with a collocated anechoic masker leads to a significant decrease in detection thresholds. Late reflections do not contribute further to unmasking because the IC does not decrease further (see Figure 8, panel a). One reason for this different outcome might be that Zurek *et al.* (2004) used for this comparison a reverberant target and masker in steady state, resulting in a decorrelation of both the noise and masker signals. In the current study, only the target sound is reverberant, resulting in potentially positive effects of reflections being more pronounced.

Zurek *et al.* (2004) also tested different absorption coefficients. For a frontal target sound source with a collocated masker, binaural detection thresholds did not differ for absorption coefficients in the range of 0.1 to 1. This result is in accordance with our findings. In the first experiment of the current study no significant difference can be found across different absorption conditions.

Braasch (2001) measured detection thresholds of a broadband noise target at different azimuth angles in the presence of a broadband noise masker in the front in a simulated reverberant room as well as in anechoic space. Detection thresholds decreased with increasing azimuth of the target sound source, in accordance with the current findings. However, thresholds differed for an anechoic versus a reverberant lateral target with a frontal noise masker, which we did not observe. Here, thresholds are not significantly different for a lateral target position comparing direct sound (anechoic) to the full RIR condition. The differences might stem from an additional detrimental effect of reflections from the reverberant noise masker used in (Braasch, 2001).

*5.2 Effects of late reflections on signal detection*

The results of the second experiment demonstrate that also isolated late reflections can improve detection thresholds: reflections arriving 75 ms after the direct sound lowered detection thresholds significantly below those for only the direct sound. These isolated late reflections decorrelate the signal and therefore increase binaural unmasking as analyzed in Figure 8 (panel b). As expected, the later the reflections arrive, the later the decorrelation of both ear signals starts. However, also reflections arriving 250 ms after the direct sound decrease the IC for the last 200 ms of the stimulus. The unmasking process for detecting a longer harmonic sound can thus benefit from the decorrelation by late reflections. For speech, however, such a benefit would be available if phonemes are voiced on the same fundamental frequency for long enough that the late reflections can still contribute energy to the harmonics. This might be the case when singing, and also for musical instrument sounds. For regular speech, the spectral speech content changes at the syllable rate of 3-4 Hz, thus preventing the add-on of similar harmonic energy from late reflections. For larger frequency changes this will limit the unmasking benefit and the reflections will interfere with the newly incoming speech sounds also in terms of the information they carry, leading to the "detrimental window" concept for late reflections which function like interfering noise. Such a segmentation in useful and detrimental energy was proposed by Bradley (1986) who showed that reflections arriving after 80 ms do not contribute to speech intelligibility in rooms. Srinivasan *et al.* (2017) measured, like most studies, speech reception thresholds and compared a full room impulse response with two truncated versions, one including only early reflections within 50 ms and one with only late reflections arriving after 50 ms. They observed lower thresholds for the condition with only early reflections compared to that with only late reflections, especially when target and noise masker were collocated in the front. Comparable findings can be found in studies by Lochner and Burger (1964) and Leclère *et al.* (2015) being all in agreement with a useful window size in the range of 50 to 80 ms. Late reflections can also contribute to speech intelligibility. Rennies *et al.* (2019) used a single late reflection 200 ms after the direct sound with the same amplitude as the direct sound but with an IPD of 180°. Listeners' speech reception thresholds decreased compared to only the direct sound if the single reflection contained binaurally favorable information (e.g. IPD of 180° ). However, a single late reflection of equal amplitude to the direct sound is likely perceived as a separate sound event.

### *5.3 Fast versus slow BMLD extraction for a binaural detection model*

Incoming reflections will cause ongoing changes of the binaural cues, affecting the unmasking of a sound source in noise as a function of time. The present article questions if such changes need to be taken into account with a dynamic model. Former detection models (Grantham and Wightman, 1979; Kollmeier and Gilkey, 1990; Holube *et al.*, 1998) have processed them with a long integration window to accout for sluggishness. Former detection models considering temporally changing signals (Braasch, 2001; Breebaart *et al.*, 2001), do not explicitly consider binaural sluggishness which is expected to influence detection thresholds. The proposed model approach in the current study tries to include and discuss the sluggish integration for detecting a reverberant signal in noise.

Recent models focus especially on speech intelligibility in reverberant listening situations (Beutelmann *et al.*, 2010; Hauth and Brand, 2018; Vicente and Lavandier, 2020). These models use two different time constants. Binaural unmasking is usually derived from a larger time frame (200 to 300 ms) whereas the monaural contribution is derived on much shorter time frames. Hauth and Brand (2018) recently extended the model from Beutelmann *et al.* (2010) by introducing a binaural temporal window of 200 ms. They extract the EC parameters within 23 ms short time block but average these parameters across 200 ms by taking the median. The averaged parameters are then used in the EC-process to derive the binaural benefit effectively on 200 ms time frames, i.e. the binaural contribution is computed from already integrated parameters. Vicente and Lavandier (2020) recently followed a related approach. They divide the input signal into 300 ms time frames to derive the binaural benefit and take sluggishness into account in one step. The better ear contribution is instead computed in 'fast' 24 ms time frames. Both models introduce a sluggish component through the integration of binaural cues in a long time window before computing the binaural benefit, assuming that the auditory system is not able to process fast changes of these cues. This differs from the approach suggested in the present paper which computes the binaural benefit on short time frames and averages afterwards. Because the BMLD computation is a non-linear operation, changing the order yields different, and, as shown here, better results.

The $BMLD_{slow}$ approach follows the concept of the above speech intelligibility models while the $BMLD_{fast}$ approach implements the idea of a sluggishness integration after the non-linear BMLD extraction stage. A fast extraction of the BMLD contribution seems to cover interaural correlation changes caused by late incoming reflections better than with an already averaged BMLD contribution over time, as visible in the lower RMSE of the predicted threshold for the second experiment. Even though the $BMLD_{fast}$ approach seems to explain the contribution of late reflections, some conditions incorporating very early reflections seem difficult. Here, the $BMLD_{fast}$ approach overestimates the binaural benefit. One reason might be that using the maximum contribution over time will use strong fluctuations caused by very early reflections in either the BMLD or the better ear SNR as a final prediction, resulting in an overestimation of the binaural benefit.

Using short evaluation time frames for BMLD contribution is also motivated in the literature which shows that the auditory system can process interaural changes in time and intensity on a short timescale (Bernstein *et al.*, 2001) of about 10 ms in certain situations. Siveke *et al.* (2008) used a noise stimulus with modulated binaural coherence and ITDs at the same time (*Phasewarp stimulus*) and contrasted it with modulation detection in monaural noise. With increasing modulation frequency, the sensitivity to detect a modulation decreases for both, the Phasewarp stimulus and monaural modulation in the same manner. They concluded that there is no indication for additional binaural sluggishness. However, the results might be affected by across-frequency processing. While interaural cues can be extracted on a short time basis, forlocalizing a tone an auditory object needs to be formed and followed which might explain the sluggish behavior observed in some studies. Building up an auditory object takes time (Bregman, 1978; Anstis and Saida, 1985; Deike *et al.*, 2012), and attaching a location to it might happen at a low rate. The conceptual advantage of a fast extraction is that fine temporal information is binaurally compared only within a short time window, reducing any requirement for a "storage".

# 6  Conclusion

The current study investigated the effect of room reflections on binaural unmasking of a low frequency harmonic complex tone in anechoic noise. The following main findings can be drawn from the current study:

- Early reflections up to 45 ms can improve binaural detection thresholds for a target in the front in the presence of a collocated, anechoic noise masker due to the decorrelation imposed on the target.
- For a lateral sound source position at 60° and a masker from the front, neither early nor late reflections contribute to further increase binaural unmasking.
- In the N0S0 condition, in the absence of early reflections and reverberation in the masker, listeners are still able to benefit from isolated late reflections up to 250 ms after the direct sound, leading to significantly decreased detection thresholds. This is because also late reflections will decorrelate the two ear signals sufficiently for a frontal target in almost diotic noise.
- A model approach computing the BMLD and monaural detection cues in short time frames (24 ms) followed by an integration to account for sluggishness and intensity integration, respectively, can predict the measured detection thresholds especially in high reverberation and isolated late reflections more accurately than when BMLDs are derived from a large time window (200 ms), which also tends to underestimate thresholds.

## Acknowledgements

This study was funded by TUM and by the Deutsche Forschungsgemeinschaft (DFG, German Research Foundation) – Projektnummer 352015383 – SFB 1330 C5. The rtSOFE system was funded by BMBF 01 GQ 1004B.

## Conflict of interest

The authors declared no conflict of interests.

# Appendix

Table A2: x, y and z coordinates of the room corners of the simulated room shown in Figure 1. The corner indexes starting in the top left corner on the foor (1-4) and were counted clockwise. Index numbers 5-8 denote the coordinates of the room ceiling also starting in the top left corder of Figure 1. The values are given in meters.

|     | x    | y     | z    |
|-----|------|-------|------|
| $S_1$ | 0    | 0     | 0    |
| $S_2$ | 0.77 | 17.49 | 0.15 |
| $S_3$ | 8.06 | 16.71 | 0.19 |
| $S_4$ | 7.24 | -0.39 | 0.31 |
| $S_5$ | 0.01 | 0.02  | 3.12 |
| $S_6$ | 0.46 | 17.14 | 2.84 |
| $S_7$ | 7.58 | 16.49 | 3.04 |
| $S_8$ | 7.01 | -0.12 | 3.35 |